Mapping excellence in the geography of science:

An approach based on Scopus data

Lutz Bornmann$, Loet Leydesdorff§, Christiane Walch-Solimena$, Christoph Ettl$

$ Max Planck Society, Hofgartenstr. 8, D-80539 Munich, Germany.

§ Amsterdam School of Communications Research, University of Amsterdam,

Kloveniersburgwal 48, NL-1012 CX Amsterdam, The Netherlands.

Corresponding author:

Lutz Bornmann, bornmann@gv.mpg.de

**Abstract**

As research becomes an ever more globalized activity, there is growing interest in national and international comparisons of standards and quality in different countries and regions. A sign for this trend is the increasing interest in rankings of universities according to their research performance, both inside but also outside the scientific environment. New methods presented in this paper, enable us to map centers of excellence around the world using programs that are freely available. Based on Scopus data, field-specific excellence can be identified and agglomerated in regions and cities where recently highly-cited papers were published. Differences in performance rates can be visualized on the map using colors and sizes of the marks.

# 1     Introduction

Bibliometric measures have led to an increased interest among science policy-makers for the identification of centers of excellence in scientific research (Danell, 2011; Frenken, Hardeman, & Hoekman, 2009). How can excellence be evaluated in terms of geographic regions and cities? In this study, we address this question and describe new methods to analyze the geographic distribution of scientific excellence. These methods allow to visualize regions (and cities within them) that are characterized by a high number of authors who published highly-cited papers. Differently from institutional rankings published thus far e.g. by SCImago Reseach Group (2010) or by Noyons et al. (2003a, 2003b), our methods allow for the spatial identification of scientific excellence on the basis of online access to the web interface of Scopus (Elsevier, Amsterdam, the Netherlands). We use programs that are freely available on the Internet and are easy to handle. The goal of our methods is to produce regional maps showing where excellent papers have emerged and where these papers have occurred frequently. Frenken et al. (2009) suggested to group these methods for mapping the geography of science under the heading 'spatial scientometrics'.

The geographic mapping of scientific papers can be distinguished from their cognitive mapping (Börner, Chen, & Boyack, 2003). In a literature overview, Frenken et al. (2009) identified some descriptive studies which investigated differences among regions or countries, respectively, in terms of their publication output and citations. Most of these studies investigated the spatial distribution of published papers in terms of output (publications) and not impact (using citations). An early study of Matthiessen and Schwarz (1999) investigated the scientific output of large European cities. These authors concluded that the north-western part of Europe showed the highest concentration of research output. The urban regions of London, Paris, Moscow, and the Dutch agglomeration (Amsterdam-Hague-Rotterdam-Utrecht) form a "super-league" of locations with high output in Western Europe.

Only a few studies focused on citation impact (instead of or in addition to publication output). Batty (2003), for example, analyzed data from the ISI's HighlyCited database (http://www.isihighlycited.com) which consisted of approximately the top 100 most cited individuals in fourteen scientific fields. Based on their first study (Matthiessen & Schwarz, 1999) which focused on publication output of large cities in Europe, Matthiessen, Schwarz, and Find (2002) followed up with an analysis of the strength, interrelations, and nodality of global research centers. In a recently published paper Matthiessen, Schwarz, and Find (2010) asked whether the next generation of hot spots in research of certain disciplines can be predicted. Focusing on the largest cities of the world and the geographic regions around these cities, these authors found "in accordance with other presented findings an extremely high and growing degree of concentration in the large centres, in combination with a series of new centres of concentration" (Matthiessen, et al., 2010, p. 1895).

Spatial bibliometrics has attracted a lot of attention in the most recent past. In *Nature News* Van Noorden (2010) discussed urban regions producing the best research and whether their success could be replicated elsewhere (Florida, 2002; Saxenian, 1996). Living Science (http://www.livingscience.ethz.ch/), created by Luis Bettencourt (Los Alamos National Laboratory in New Mexico) and collaborators under Dirk Helbing at the ETH Zurich track where arXiv papers are published in real time. In general it pays off for the sciences within a country to identify (by visualization methods) and expand regional centers of excellence (with specific financial support). As a rule, there is a high probability of co-operation between scientists working at a short physical distance (Katz, 1994). Synergies between ideas and direct face-to-face communication between scientists are considered as major factors of productivity (Matthiessen, et al., 2002; Wagner, 2008).

## 2 Methods

Percentile citation impact classes are well suited for identifying the highly cited papers in a scientific field (Bornmann & Mutz, 2011). However, in evaluative bibliometrics there is uncertainty regarding within which percentile rank a paper would have to be considered as highly cited. According to Tijssen, Visser, and van Leeuwen (2002) and Tijssen and van Leeuwen (2006), highly cited papers are those among the top 10% of the most cited papers – that is, papers in or greater than the 90$^{th}$ percentile of a field (cf. Lewison, Thornicroft, Szmukler, & Tansella, 2007). In the Essential Science Indicators (ESI) Thomson Reuters classifies as highly cited papers those that belong to the top 1% of papers worldwide (papers in or greater than the 99$^{th}$ percentile), taking into account the field and year of publication. According to the National Science Board (2010) top-performance or highly cited papers are those in the top 1% also. In this study we follow the classification of Thomson Reuters and the National Science Board (2010), because the methods proposed here can process no more than 2000 papers due to a systems limit of the Scopus data base. Scopus is used as data base for this study since this is currently the only database in which one can select all papers published in a broader field (e.g., physics & astronomy). A study using the top 10% of the most highly cited papers in a scientific field would exceed this limit in most of the fields. We focus on the top 1% of papers published in 2007 with a fixed citation window of three to four years (from 2007 up to the date of research, at the end of 2010 and in March/April 2011, respectively).

In the following, the procedure to map the excellent papers (more precisely: the cities of the authors having published the top 1% most highly cited papers) in a certain field is described. The procedure will be explained for the field of "neuroscience". With the search string "subjarea(neur) and pubyear is 2007 and doctype(ar)" in the advanced search field of Scopus all papers with the document type "article" are retrieved which were published in 2007 within the Scopus journal set of "neuroscience." For the date of April 5, 2011 this search resulted in 40,082 papers (see

Table 1). The search was restricted to articles (as document types) since (1) the method proposed here is intended to identify excellence at the research front and (2) different document types have different expected citation rates, possibly resulting in non-comparable datasets.

Table 1. Data used for the maps

| Field | Published articles in 2007 | Citation limit for the top 1% most highly cited articles | Number of articles belonging to the top 1% most highly cited articles |
|---|---|---|---|
| Neuroscience | 40,082 | 69 | 407 |
| Physics and astronomy | 146,081 | 44 | 1,501 |
| Social sciences | 76,534 | 27 | 759 |
| *Nature* and *Science* articles (Scopus) | 1,754 | | |
| *Nature* and *Science* articles (Web of Science) | 1,604 | | |

By sorting the search results by citation counts in decreasing order (citation window: from 2007 to the date of search), the 1% of papers at the top of the Scopus list can be marked. At the date of search, 407 papers with at least 69 citations each (gathered between 2007 and the date of search) were marked as the list of the top 1% neuroscience papers. Although one percent of 40,082 is 401, we included ranks which were tied at this 1% level and thus retrieved 407 records. The selected documents are exported by choosing the export format "Comma separated file, .csv (e.g. Excel)" and the output "Specify fields to be Exported." Only the field "Affiliations" is selected and exported. The download in the .csv format can be processed with the program scopcity.exe. This and the programs mentioned below including the respective user instructions can be retrieved from

http://www.leydesdorff.net/mapping_excellence/index.htm (see here also Leydesdorff & Persson, 2010).

The programs and the .csv file must be stored in the same folder. scopcity.exe creates, among other files, cities.txt. This file contains all city entries of the top-1% papers downloaded from Scopus (see here Costas & Iribarren-Maestro, 2007). If there is more than one co-author of a publication with an identical address, this leads to a single address (or a single city occurrence) in cities.txt. If the scientists are affiliated with different departments within the same institution, this leads to two addresses or two city occurrences, respectively. The content of cities.txt can then be copied-and-pasted into the GPS encoder at http://www.gpsvisualizer.com/geocoder/. Since no more than 1000 entries can be processed by the encoder, more than 1000 entries in cities.txt must be entered into the encoder in subsequent steps. Our experiences suggest that the better encoding results are obtained if Google instead of Yahoo! is chosen as the source of the coordinates. The best procedure is to use both Google and Yahoo! and to check whether they give more or less consistent results.

After saving the results in the output window of the geo-encoder as the DOS text file named "geo.txt" this data serves as input for cities2.exe. It is mandatory that geo.txt contains all entries from cities.txt with the additional geo data—in precisely the same order because this order is used for matching the two files. Running Cities2.exe produces a number of output files in various formats within the folder (Leydesdorff & Persson, 2010). After cities2.exe is finished, cities3.exe should be used as a final step in the analysis. This routine algorithm aggregates similar city names with somewhat different geocodes because of different institutional addresses. This program needs the following user input: "Do you wish to use the information for a detailed map (1; r=0.01) or rather for a mapping of metropolitan areas (3; r=0.3)? Option (2) provides an in-between option (r=0.1)." Option 1 should be chosen to test a map initially. The test procedure is explained later on. Otherwise options 2 or 3 can be chosen (for the maps presented in this paper option 2 were chosen). These routines merge locations around a city with approximately the same geo-coordinates, but for example different postcodes or zip-codes. Although options 2 and 3 are also intended as an error-correction mechanism to reduce problems like "Zurich" and "Zrich" or "Munich" and "Munchen," not all problems could be solved using Scopus data.

Furthermore, cities3.exe makes sure that within each map, the circles are coloured according to the number of authors of excellent papers in each city. Note that the colouring is dependent on the options (1, 2 or 3) for the aggregation used in cities3.exe. This means that circles which are produced on the basis of the same data may appear in different colours using options 1, 2 or 3. The colorization of the circles supports the visualisation of the different numbers by different circle radii. The radii are proportional to the logarithm of the number of top-1% cited papers. With this feature the viewer of a map may realise those cities with the highest (and lowest) numbers more readily.

We used the percentile rank approach proposed by Bornmann and Mutz (2011) to colorize the circles. The percentiles were computed as follows: First, the numbers of authors (of excellent papers) $X_i$ for the $i^{th}$ city within n cities (of one map) were ranked in decreasing order

$$X_1 \geq X_2 \geq ... \geq X_n,$$

where $X_1$ ($X_n$) denotes the number of papers associated with city names or, in other words, with the largest (lowest) number of author addresses. Secondly, for each city, a percentile rank based on this distribution is assigned. If, for example, a single city acquires 50 papers whereas 90% of the other cities have 49 papers or less, then this particular city would be in the $90^{th}$ percentile.

All cities (circles on a single map) are categorized into six percentile rank classes and coloured accordingly:

Red circle: top 1% (cities with a percentile equal to or greater than the $99^{th}$ percentile),

Fuchsia circle: $95^{th}$ – $99^{th}$ (cities within the $[95^{th}; 99^{th}[$ percentile interval),

Pink circle: $90^{th}$ – $95^{th}$ (cities within the $[90^{th}; 95^{th}[$ percentile interval),

Orange circle: $75^{th}$ – $90^{th}$ (cities within the $[75^{th}; 90^{th}[$ percentile interval),

Cyan circle: $50^{th}$ – $75^{th}$ (cities within the $[50^{th}; 75^{th}[$ percentile interval),

Blue circle: bottom 50% (cities with a percentile less than the $50^{th}$ percentile).

"ucities.txt," that is, the output file of cities3.exe, can be uploaded into the GPS Visualizer at http://www.gpsvisualizer.com/map_input?form=data. The Web page offers a number of parameters to visualize the data in ucities.txt. The following parameters should be set: (a) change "waypoints" into "default;" (b) as background map choose "Google street map" and as background map opacity "80%;" (c) change "colorize using this field" into "custom field;" (d) change "resize using this field" into "custom field" and write "n" into "custom resizing field;" and (e) choose at "show point names" option "No."

When the GPS data has been processed, the Google map is displayed in a small frame, but it is also temporarily available to view on the full screen. The map shows the regional distribution of the authors of highly cited papers. The background map's opacity can be adjusted or another layout as available in Google or Yahoo! can be chosen. With the instruments visualized on the left side of the map it is possible to zoom into the map. First, the the GPS Visualizer shows the global map. For the maps presented in the following we zoomed into Europe in order to generate comparable maps for different publication sets. Similarly, other regional foci can be chosen. To determine the number of papers for a specific circle, one can click on the respective circle. Maps generated in this way can be copied to other programs (like Microsoft Word) by using programs utilized for screen shots (e.g., Hardcopy). If one uses the download instead of the view command shown in the Google Maps output page, a html-coded page is saved that includes the data of ucities.txt. Opening this page within a browser will regenerate the respective Google Map. Google Maps offers free APIs for using these maps on one's own website.

There are several problems inherent to the approach proposed here. The user should always be aware of these limitations when the approach is applied:

1) There are circles (as a rule blue circles) on the maps that are not at the right position. In the various routines, we tried to avoid these misallocations, but they could not be completely resolved. The misallocations do have different sources: errors in the Scopus data or erroneous coordinates provided by the Geocoder. For example, there are two papers (Pakhlova et al., 2007; Starič et al., 2007) in the physics data set with the following address: "Swiss Federal Institute of Technology of Lausanne, EPFL, Lausanne, 61801." The appendix "61801" is mistaken. This results in faulty assignment of coordinates by the Geocoder (independent of the source of the coordinates, Google or Yahoo!). A complete data cleaning procedure cannot be provided by our approach automatically, but this can be done by the user manually.

2) For some locations, the GPS encoder using Google or Yahoo! as source for the coordinates are not able to find the coordinates and adds "0,0" (e.g., "Ben-gurion University Of The Negev, Israel" produces this using Google as source). Yahoo! is reported to be more precise with Asian addresses; however in our hands the Google search engine produced the most reliably results. One can also try to fill out the gaps with "0,0" using the other search engine. Addresses starting with „0,0" as coordinates are deleted during the processing of data (in cities2.exe) and they are consequently not visualized on the maps.

3) The methods as described above do not allow for the identification of the research institutions on the map where the authors of the excellent papers are located.

4) Since it can be assumed for certain fields (e.g., the life sciences) that authors at specific positions in the list of authors (e.g., the first or the last authors) have made particularly significant contributions to a publication it would be interesting to include in the analysis a restricted set of authors (e.g. only the first authors). However, this is not yet possible with our approach.

5) High numbers of publications visualized on the map for one single city might be due to the following two effects: (a) Many scientists located in that city (i.e., scientists at different institutions or departments within one institution) produced at least one excellent paper or (b) one or only a few scientists located in this city produced many influential papers. Assuming cities as units of analysis, one is not able to distinguish between these two interpretations.

6) Because of a systems limit in Scopus only 2,000 papers can be retrieved and downloaded from the Web interface. Thus, fields with more than 2,000 papers among the top 1% – that is, above a total of 200,000 – cannot be visualized using the approach presented here. One may in this case wish to set higher thresholds and study, for example, the 1‰ most highly cited papers.

Each map produced by applying our approach should be carefully checked. This can be done using the following two procedures: 1) A map is produced through option 1 in cities3.exe and is visualized showing point names on the map (see above). Using the point name (e.g., Stuttgart) the number of highly cited papers for a particular circle (e.g., in the field physics & astronomy") can be checked in Scopus with the search string "subjarea(phys) and pubyear is 2007 and doctype(ar) and AFFILCITY(stuttgart) and AFFILCOUNTRY(Germany)." If the search results in Scopus are ordered by times cited, the number of papers above the citation threshold (established for that particular field) should be more or less equal to the number of papers shown for that circle. Cities with the same name located in the same countries (e.g., Cambridge, MA, USA, and Cambridge, ID, USA) should be checked primarily; but also multiple cities with the same name in different countries (e.g., Cambridge, UK, and Cambridge, MA, USA). However, our methods are hitherto based on integer counting. If two addresses in the same city are provided on one document (such as in the case of inter-institutional collaborations), the same city name is counted twice and the retrieval using Scopus will consequently be lower than the numbers indicated on the maps. 2) Using the point names or city names, respectively, on the map it should be checked for some circles whether they are located on the right positions. As described above the Geocoder adds sometimes faulty coordinates (independently of the source of the coordinates, that is, Google or Yahoo!).

Only if both procedures do not bring about conspicuous errors, the map can safely be used and uploaded as a result.

## 3 Results

To demonstrate the proposed method we produced three field-specific maps.

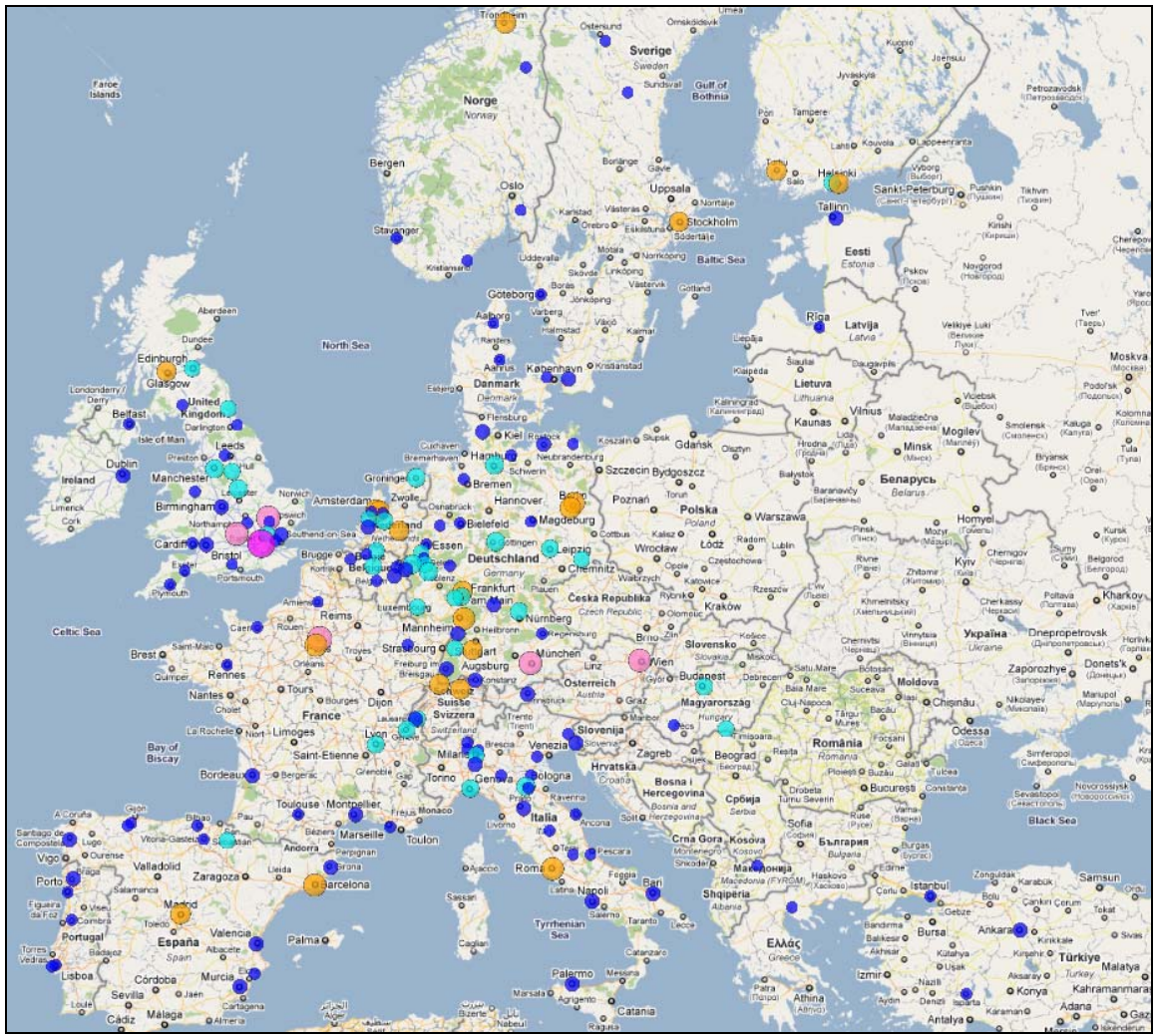

Figure 1. Locations of authors in Europe having published highly cited neuroscience papers in 2007 (this figure appears in colour on the Web (PDF and HTML of this paper), but is not reproduced in colour in the printed version). See also:
http://www.leydesdorff.net/mapping_excellence/figure1.htm

Figure 1 shows the location of authors in Europe who published highly cited papers in neuroscience. The map is based on the top 1% of articles published in 2007 in a journal of the Scopus journal set neuroscience (see above and Table 1). On the map it is indicated in which regions the authors are located (the circles with different colors on the map) and the frequency of occurrences (author addresses of the papers) per location: (1) the radius of the circles is proportionate to the logarithm of the frequency; (2) the colors of the circles correspond with the percentile rank classes of the cities indicated.

For example, Bari in Italy is classified on this map as belonging to the bottom 50%. For Vienna in Austria, the larger pink circle reveals this city within the [90$^{th}$; 95$^{th}$[ percentile interval; Vienna belongs to the top 10% of the cities worldwide locating the most authors of excellent neuroscience papers. If one focuses on geographic regions with a high concentration of excellent papers, these regions are as a rule situated around larger cities (e.g, Vienna, Paris, Munich or London). According to Matthiessen et al. (2010) "cities are almost always also centres for a hinterland, which they more or less dominate" (p. 1880). On the map, a couple of regions with a higher density of circles around a centre are visible; one region is very salient: London – Cambridge – Oxford.

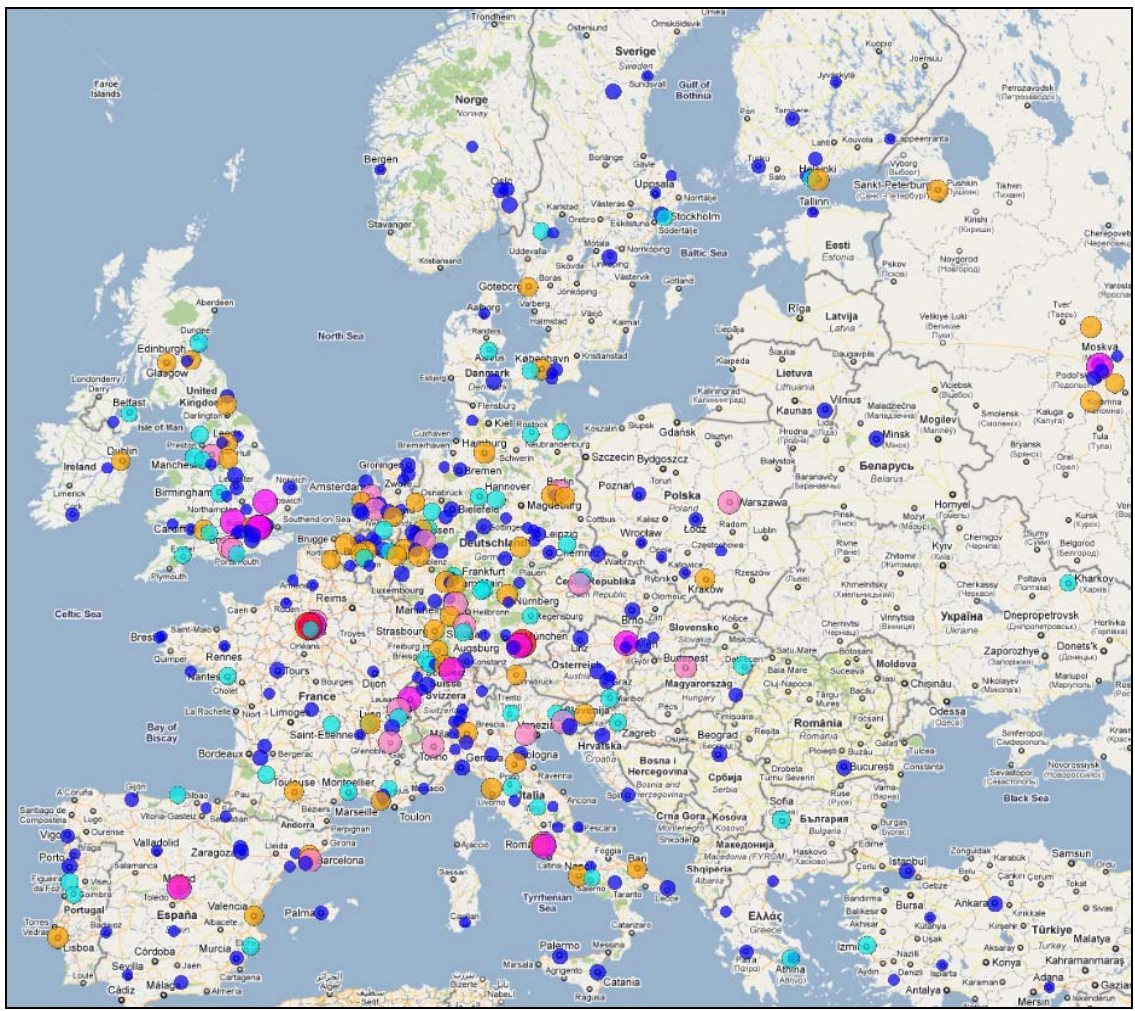

Figure 2. Locations of authors in Europe having published highly cited physics and astronomy papers in 2007 (this figure appears in colour on the Web (PDF and HTML of this paper), but is not reproduced in colour in the printed version). See also:
http://www.leydesdorff.net/mapping_excellence/figure2.htm

Figure 2 shows the corresponding map for physics and astronomy (Table 1). In 2007, 146,081 articles were published in this journal set worldwide; the top 1% are those 1,501 papers which received at least 44 citations each between 2007 and the date of research (March 29, 2011). Here again, we see a higher density of circles around London – Cambridge – Oxford with three fuchsia circles. Paris and Garching (near Munich) belong to the top 1% of the cities worldwide with the highest numbers of authors of excellent papers.

Since physics & astronomy and neuroscience are natural science fields it is interesting to see how the European map changes in the case of the social sciences. Figure 3 shows the map for the authors of 759 top cited papers (selected from 76,534 articles published in 2007) which received at least 27 citations each between 2007 and the date of research (April 5, 2011) (see

Table 1). It is clearly visible that German cities are significantly less frequently locations of highly cited authors in the social sciences than in physics & astronomy. For Europe, there are two fuchsia circles on the map: Zurich and London are among the top 5% of the cities worldwide with the highest numbers of excellent papers in this field.

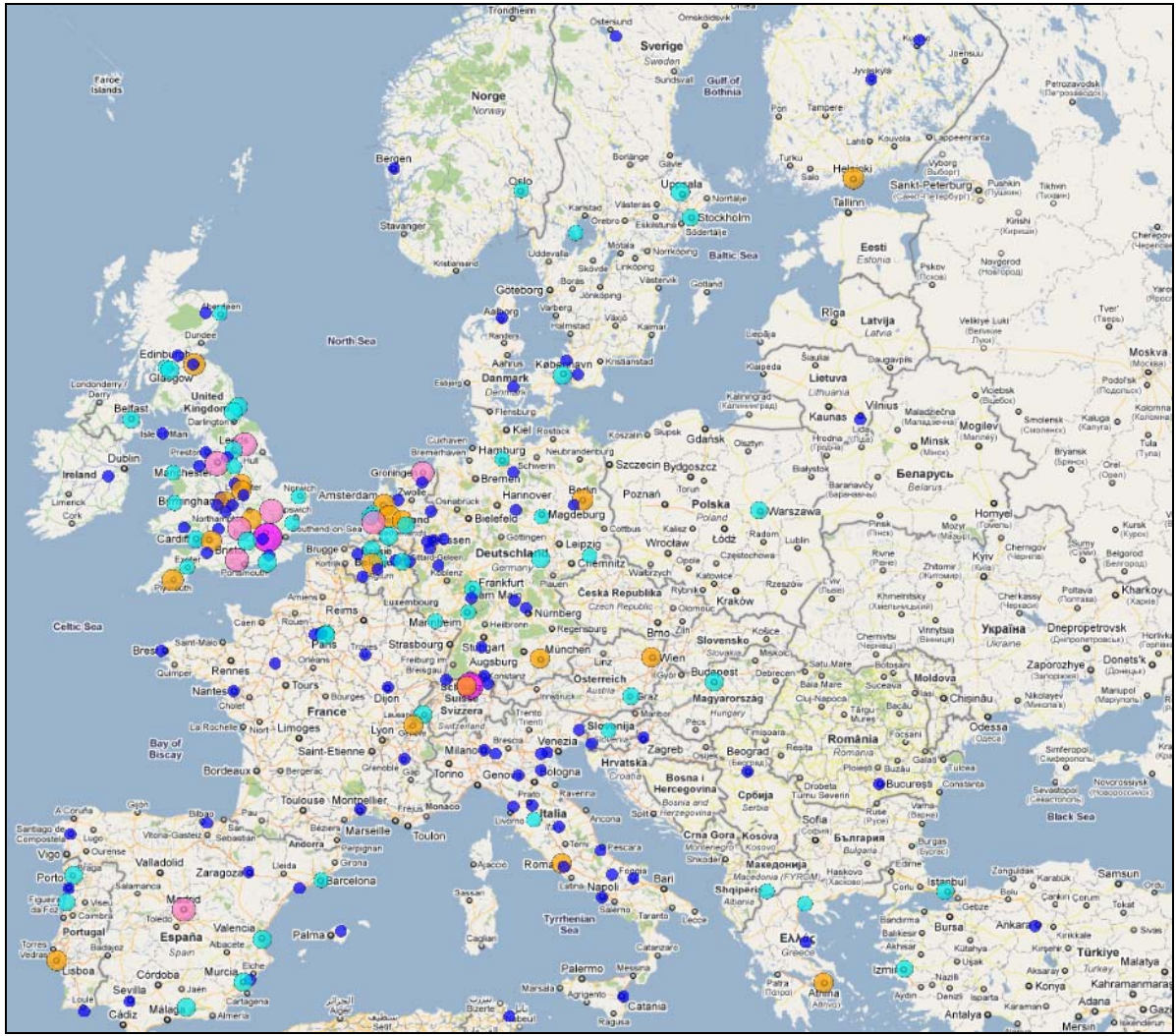

Figure 3. Locations of authors in Europe having published highly cited social sciences papers in 2007 (this figure appears in colour on the Web (PDF and HTML of this paper), but is not reproduced in colour in the printed version). See also:
http://www.leydesdorff.net/mapping_excellence/figure3.htm

As a final step in the analyses for this paper we left the focus on specific fields and performed a fields-overlapping analysis of excellent papers. For this, we downloaded from Scopus the bibliographic data of all articles published in 2007 in the high-impact journals *Nature* and *Science* (

Table 1). This analysis is inspired by the maps of Luis Bettencourt and Jasleen Kaur (Indiana University in Bloomington) published on www.nature.com/news/specials/cities/best-

cities.html. The authors analyzed city addresses appearing in *Science, Nature*, and *Proceedings of the National Academy of Sciences* in 1989, 1999, and 2009.

We downloaded the bibliographic data of all 1,754 articles published in 2007 (the date of this download was April 6, 2011). For comparison, we repeated the same analysis with Web of Science (WoS, Thomson Reuters) data (cf. Leydesdorff & Persson, 2010). From WoS we downloaded 1,604 *Nature* and *Science* articles that were published in 2007 (the date of the download from the WoS was October 27, 2010). The difference between the WoS and Scopus in the numbers of articles can probably be accounted for by the fact that publications are sometimes differently categorized as a certain document type (here: article) in both data bases (Leydesdorff & Opthof, 2011).

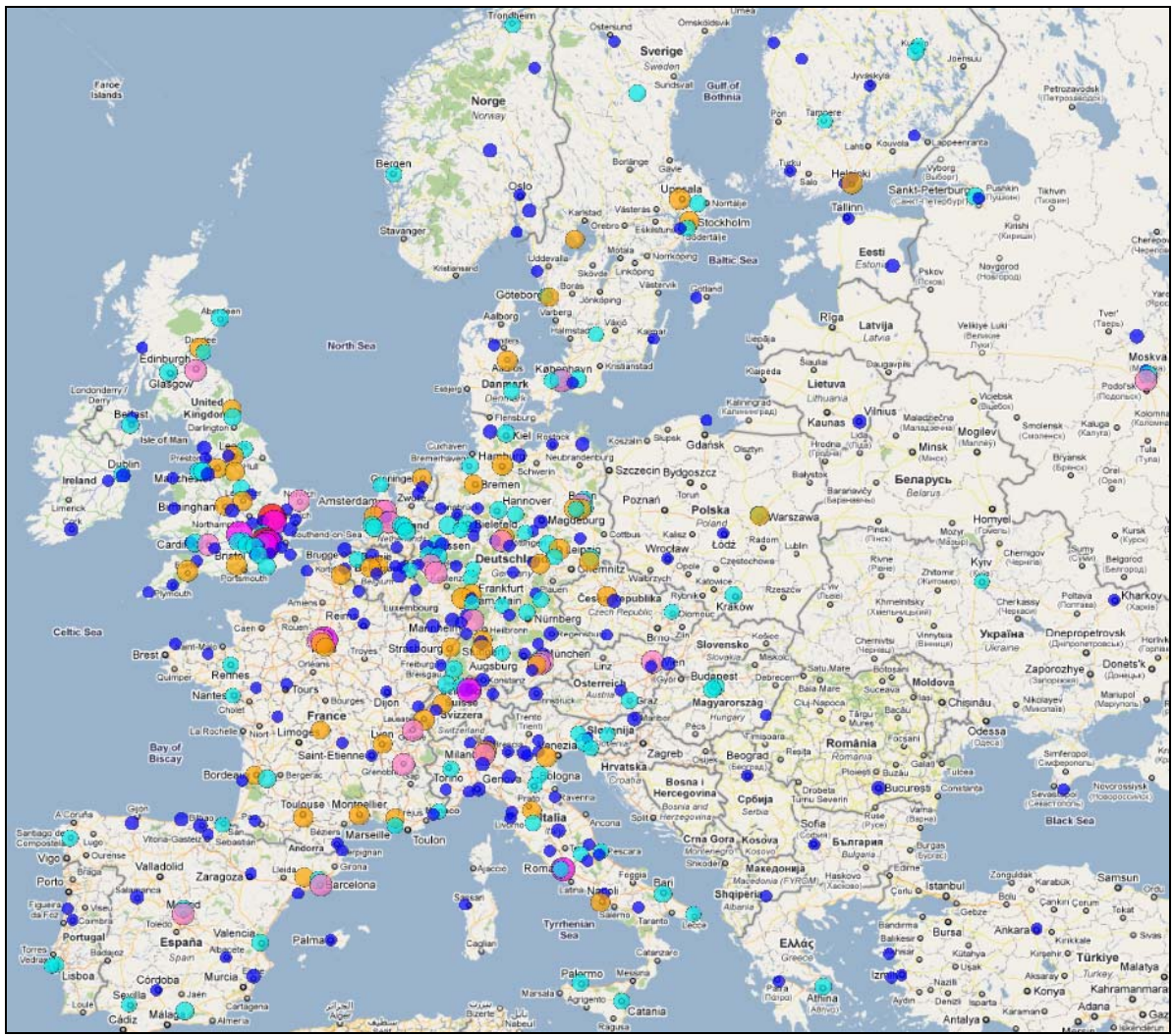

Figure 4. Locations of authors in Europe having published *Nature* or *Science* articles in 2007 (searched in Scopus; this figure appears in colour on the Web (PDF and HTML of this paper),

but is not reproduced in colour in the printed version). See also:
http://www.leydesdorff.net/mapping_excellence/figure4.htm

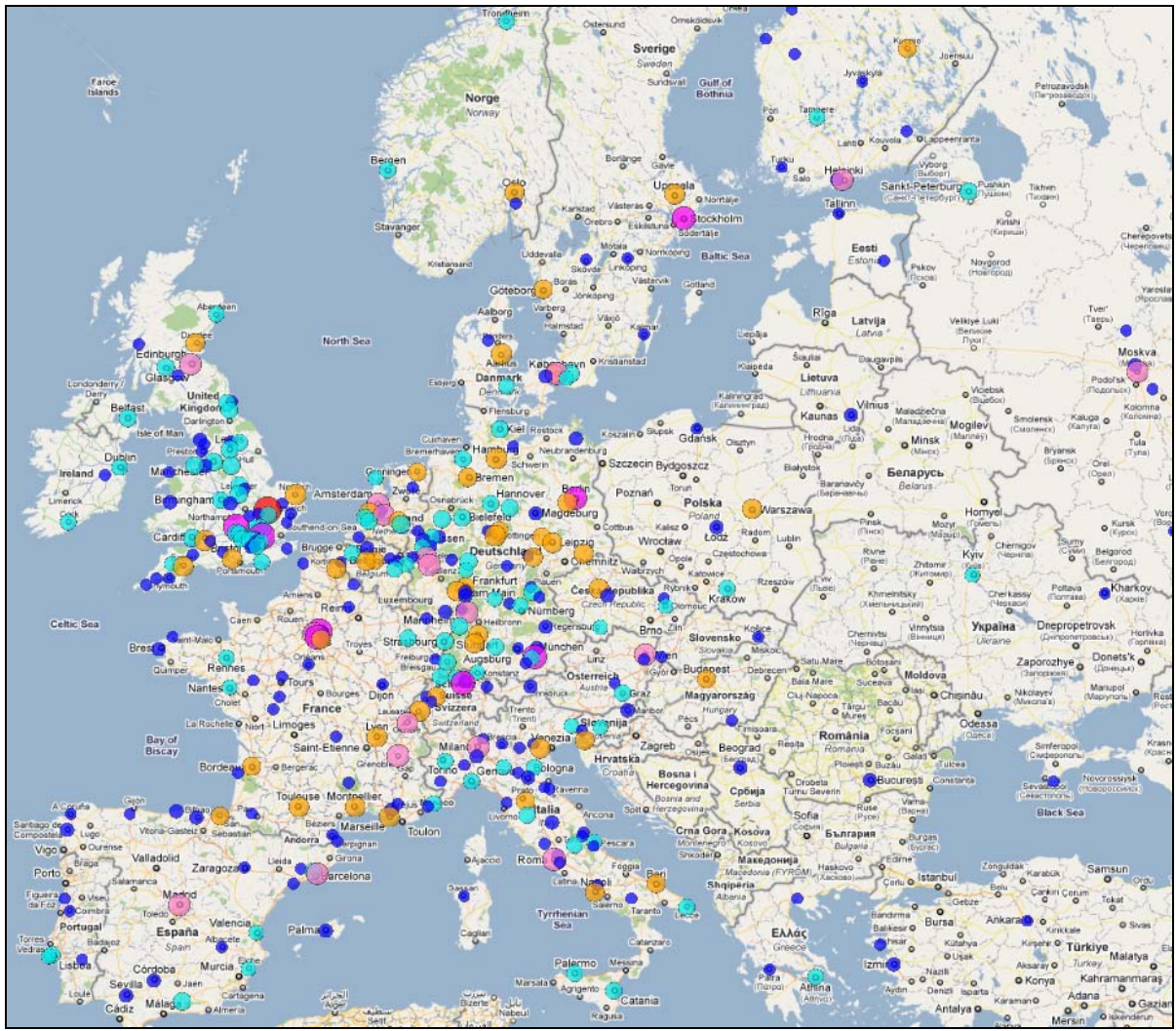

Figure 5. Locations of authors in Europe having published *Nature* or *Science* articles in 2007 (searched in Web of Science; this figure appears in colour on the Web (PDF and HTML of this paper), but is not reproduced in colour in the printed version). See also:
http://www.leydesdorff.net/mapping_excellence/figure5.htm

The maps based on the Scopus and WoS data are shown in Figure 4 and Figure 5, respectively. The results are similar. In agreement with the field-specific maps presented above, higher densities of circles in the neighborhoods of London – Cambridge – Oxford and around large cities (e.g., Paris) are visible in both presentations. The differences between the maps based on the two data bases are due to inconsistencies in the Scopus data; the data format is more consistent in the WoS. In WoS, the addresses are more standardized under certain rules. As a rule this may lead to more and other colored circles around a single city (see, e.g., Berlin) using the Scopus data while this is unambiguous using WoS data.

## 4    Discussion

As research becomes ever more a globalized activity, there is growing interest in national and international comparisons of standards and quality in different countries and regions. A sign for this trend is an increasing interest in rankings of universities according to their research prowess, both inside but also outside the scientific environment. The methods presented in this paper allow for an analysis revealing centers of excellence around the world using programs that are freely available. Based on Scopus data, field-specific excellence can be identified in regions and cities where highly-cited papers were published. The programs used for our approach are constantly refined. For that we appreciate feedback of users on improvements.

As we could show the spatial concentration of scientific activity in Europe is remarkable. The analyses point out, e.g., that the region London – Cambridge – Oxford is characterized by a high number of excellent research output – measured by top-cited papers or papers published in *Nature* or *Science*. Against the backdrop of this finding and the results of other studies (Frenken, et al., 2009) we propose to introduce a name for this spatial concentration phenomenon in science: the reverse N-effect. The formulation of the N-effect goes back to Garcia and Tor (2009). Their socio-psychological research revealed that more competitors performing the same task produce less competition because of motivation-losses.

In science, one can assume a reverse N-effect: More competitors (here: prolific scientists) working within the same region produce better results. As this paper shows, the better result may consist of a higher output of highly-cited papers; the findings of Lee, Brownstein, Mills, and Kohane (2010) provide evidence for the role of such a reverse N-effect as a predictor of the citation impact of collaborations. Due to a systematic spatial bias in the interaction prolific researchers thus favor cooperation with other prolific researchers located in physical closeness. According to the results of Acosta, Coronado, Ferrándiz, and León (2011) "geographical distance and contiguity are both relevant variables in explaining academic scientific collaboration between regions. The negative sign of the first variable indicates that collaboration decreases with distance, while the positive sign of the second variable shows that bordered regions explain their scientific collaborative behaviour" (pp. 71-72).

According to Frenken et al. (2009) "at least three mechanisms may explain why interactions in science are spatially biased towards physically proximate actors. First, serendipitous encounters are more likely when two actors are in close vicinity of each other. Second, the need for face-to-face interaction when engaging in interactions comes at a cost, which increases as a function of travel time. Third, 'the rules of the game' that matter for scientific knowledge production (e.g. funding, labour market regimes, intellectual property right regimes, languages) are spatially differentiated and constrain interaction between institutional frameworks, in particular, between nation-states" (p. 224).

With regard to the reverse N-effect in science it could be interesting to study whether an accumulation effect over time is visible at certain regions worldwide. Do regions where more excellent papers are published produce increasingly excellent papers? The literature research of Frenken et al. (2009) found only two such studies. Both studies point out agglomeration advantages to be present though the evidence is preliminary.

Despite the advantages of the approach to map excellence in science proposed in this study, we recognize the limitations inherent to bibliometric data. (1) Publications are only one among several types of scientific activities. (2) It is not guaranteed that the addresses listed on the publication reflect the locations where the reported research was conducted. (3) The handling of multiple authorships differs in scientific fields. (4) No standard technique exists for the subject classification of articles (Bornmann & Daniel, 2008; Bornmann, Mutz, Neuhaus, & Daniel, 2008; Leydesdorff & Rafols, 2009).

Nevertheless the proposed method allows for the generation of maps providing novel insights into locations of particular strength in a given field of science. Furthermore, this tool could be used in the future to assess the dynamics of knowledge-generation when applied to compare maps over longer periods of time.